\documentclass[pre,twocolumn,superscriptaddress]{revtex4}  

\usepackage[pdftex]{graphicx}
\usepackage{amsmath,amsfonts,amssymb}
\usepackage{morefloats}
\usepackage{tabularx}

\usepackage{hyperref}
\usepackage{xcolor} 

\definecolor{bluemoi}{rgb}{0.25,0.50 ,0.75} 

\hypersetup{
  bookmarksopen=false,
  pdftitle=" ",
  pdfauthor=" ", 
  pdfsubject=" ", 
  pdftoolbar=true, 
  pdfmenubar=true, 
  pdfhighlight=/O, 
  colorlinks=true, 
  pdfpagemode=UseNone, 
  pdfpagelayout=SinglePage, 
  pdffitwindow=true, 
  linkcolor=bluemoi, 
  citecolor=bluemoi, 
  urlcolor=bluemoi 
}

\makeatletter
\renewcommand{\figurename}{Figure}
\makeatother
\makeatletter
\renewcommand{\fnum@figure}{\small\textbf{\figurename~\thefigure}}
\makeatother

\begin{document}

\title{Tweets on the road}

\author{Maxime Lenormand}\affiliation{Instituto de F\'isica Interdisciplinar y Sistemas Complejos IFISC (CSIC-UIB), Campus UIB, 07122 Palma de Mallorca, Spain}

\author{Ant{\`o}nia Tugores}\affiliation{Instituto de F\'isica Interdisciplinar y Sistemas Complejos IFISC (CSIC-UIB), Campus UIB, 07122 Palma de Mallorca, Spain}

\author{Pere Colet}\affiliation{Instituto de F\'isica Interdisciplinar y Sistemas Complejos IFISC (CSIC-UIB), Campus UIB, 07122 Palma de Mallorca, Spain}

\author{Jos\'e J. Ramasco}\affiliation{Instituto de F\'isica Interdisciplinar y Sistemas Complejos IFISC (CSIC-UIB), Campus UIB, 07122 Palma de Mallorca, Spain}

\begin{abstract} 
The pervasiveness of mobile devices, which is increasing daily, is generating a vast amount of geo-located data allowing us to gain further insights into human behaviors. In particular, this new technology enables users to communicate through mobile social media applications, such as Twitter, anytime and anywhere. Thus, geo-located tweets offer the possibility to carry out in-depth studies on human mobility. In this paper, we study the use of Twitter in transportation by identifying tweets posted from roads and rails in Europe between September 2012 and November 2013. We compute the percentage of highway and railway segments covered by tweets in $39$ countries. The coverages are very different from country to country and their variability can be partially explained by differences in Twitter penetration rates. Still, some of these differences might be related to cultural factors regarding mobility habits and interacting socially online. Analyzing particular road sectors, our results show a positive correlation between the number of tweets on the road and the Average Annual Daily Traffic on highways in France and in the UK. Transport modality can be studied with these data as well, for which we discover very heterogeneous usage patterns across the continent. 
\end{abstract}

\maketitle

\section{INTRODUCTION}

An increasing number of geo-located data are generated everyday through mobile devices. This information allows for a better characterization of social interactions and human mobility patterns \cite{Watts2007,Vespignani2009}. Indeed, several data sets coming from different sources have been analyzed during the last few years. Some examples include cell phone records \cite{Onnela2007,Eagle2009,Gonzalez2008,Song2010,Phithakkitnukoon2012,Wang2009,Ratti2006,Reades2007,Soto2011,Frias2012,Isaacman2012,Toole2014,Pei2013,Louail2014}, credit card use information \cite{Hasan2012}, GPS data from devices installed in cars \cite{Gallotti2012,Furletti2013}, geolocated tweets \cite{Mocanu2013,Bailon2011,Hawelka2013,Lenormand2014} or Foursquare data \cite{Noulas2012}. This information led to notable insights in human mobility at individual level \cite{Gonzalez2008,Hawelka2013}, but it makes also possible to introduce new methods to extract origin-destination tables at a more aggregated scale \cite{Phithakkitnukoon2012,Isaacman2012,Lenormand2014}, to study the structure of cities \cite{Louail2014} and even to determine land use patterns \cite{Soto2011,Frias2012,Pei2013,Lenormand2014}. 

In this work, we analyze a Twitter database containing over $5$ million geo-located tweets from $39$ European countries with the aim of exploring the use of Twitter in transport networks. Two types of transportation systems are considered across the continent: highways and trains. Tweets on the road and on the rail between September 2012 and November 2013 have been identified and the coverage of the total transportation system is analyzed country by country. Differences between countries rise due to the different adoption or penetration rates of geo-located Twitter technology. However, our results show that the penetration rate is not able to explain the full picture regarding differences across counties that may be related to the cultural diversity at play. The paper is structured as follows. In the first section, the datasets are described and the method used to identify tweets on highways and railways is outlined. In the second section, we present the results starting by general features about the Twitter database and then comparing different European countries by their percentage of highway and railway covered by the tweets. Finally, the number of tweets on the road is compared with the Average Annual Daily Traffic (AADT) in France and in the United Kingdom to assess its capacity as a proxy to measure traffic loads.

\begin{figure*}
\begin{center}
\includegraphics[scale=1.2]{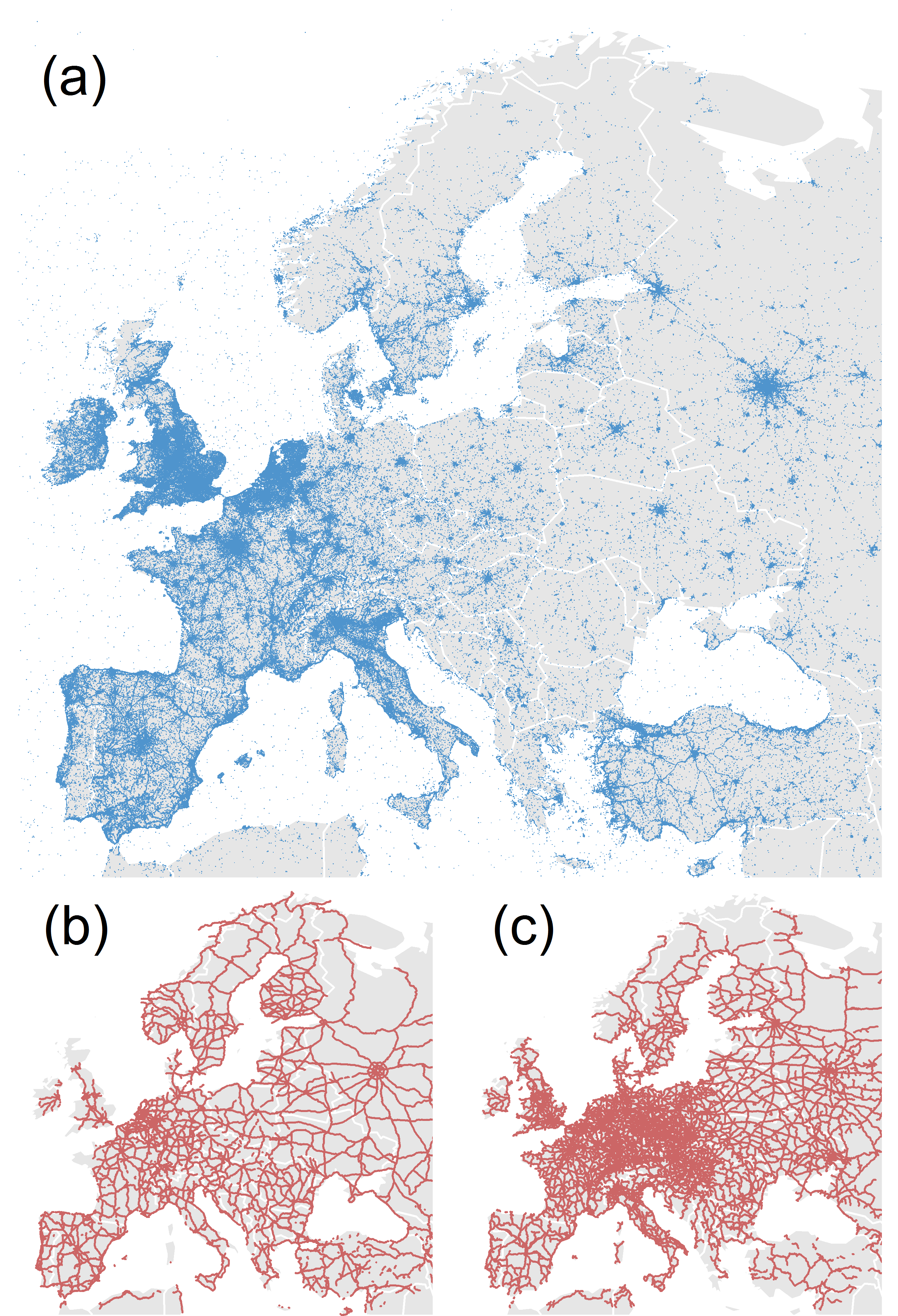}
\caption{(a) Geo-located tweets on a map. (b) Highway network. (c) Railway network. \label{map}}
\end{center}
\end{figure*}

\begin{figure*}
\begin{center}
\includegraphics[width=\linewidth]{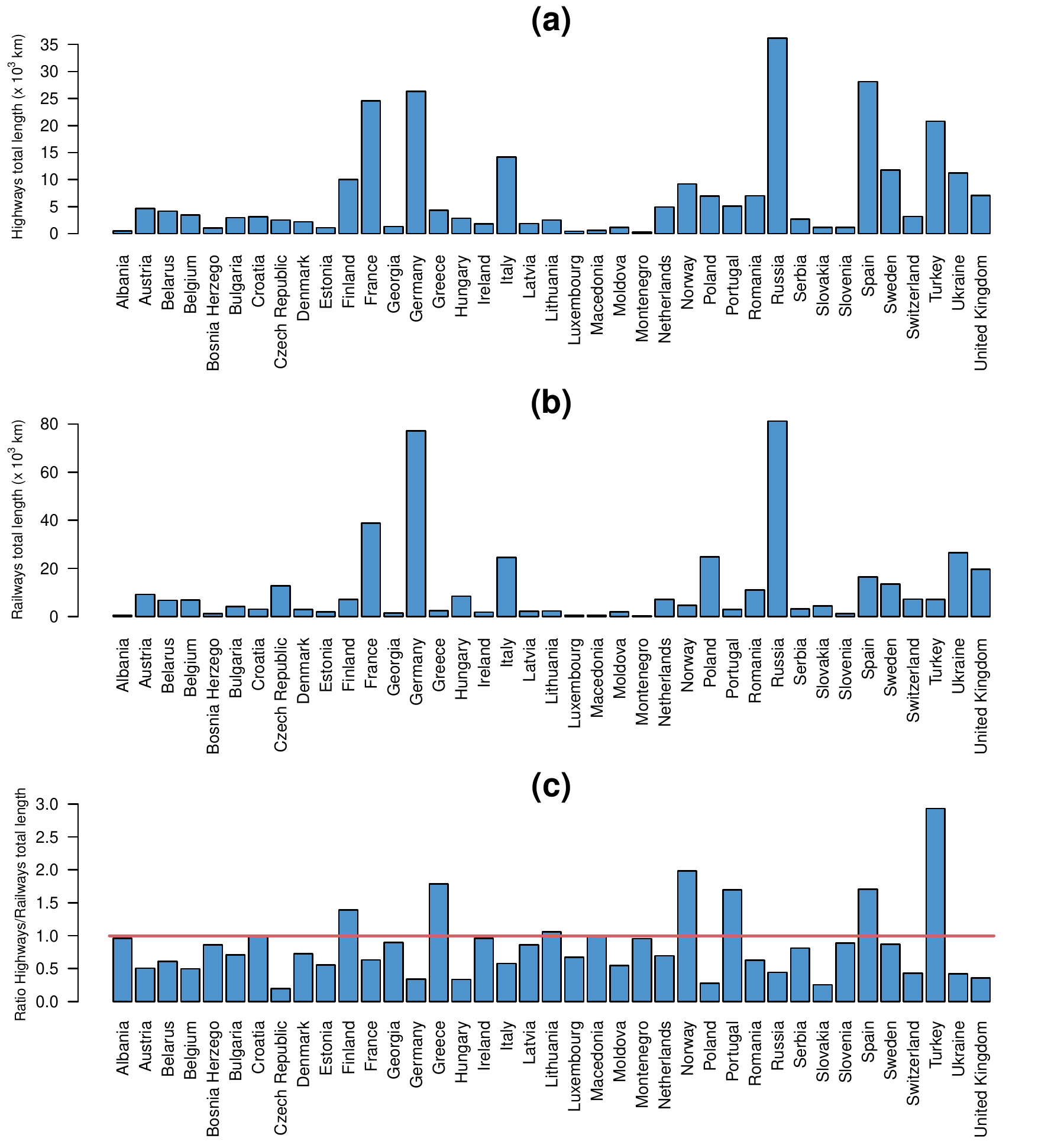}
\caption{(a) Highways total length by European countries. (b) Railways total length by European countries. (c) Ratio  between the total length of highways and that of railways by several European countries. The red line marks the unit ratio. \label{RankRoad}}
\end{center}
\end{figure*}
 
\section{MATERIALS AND METHODS}

\subsection{Datasets}

The dataset comprehends $5,219,539$ geo-located tweets across Europe emitted by $1,477,263$ Twitter users in the period going from September $2012$ to November $2013$. The data was gathered through the general data streaming with the Twitter API \cite{API}. It is worth noting that the tweets are not uniformly distributed, see Figure \ref{map}a. Countries of western Europe seem to be well represented, whereas countries of Eastern Europe are clearly under-represented (except for Turkey and Russia).

\begin{figure*}
\begin{center}
\includegraphics[width=\linewidth]{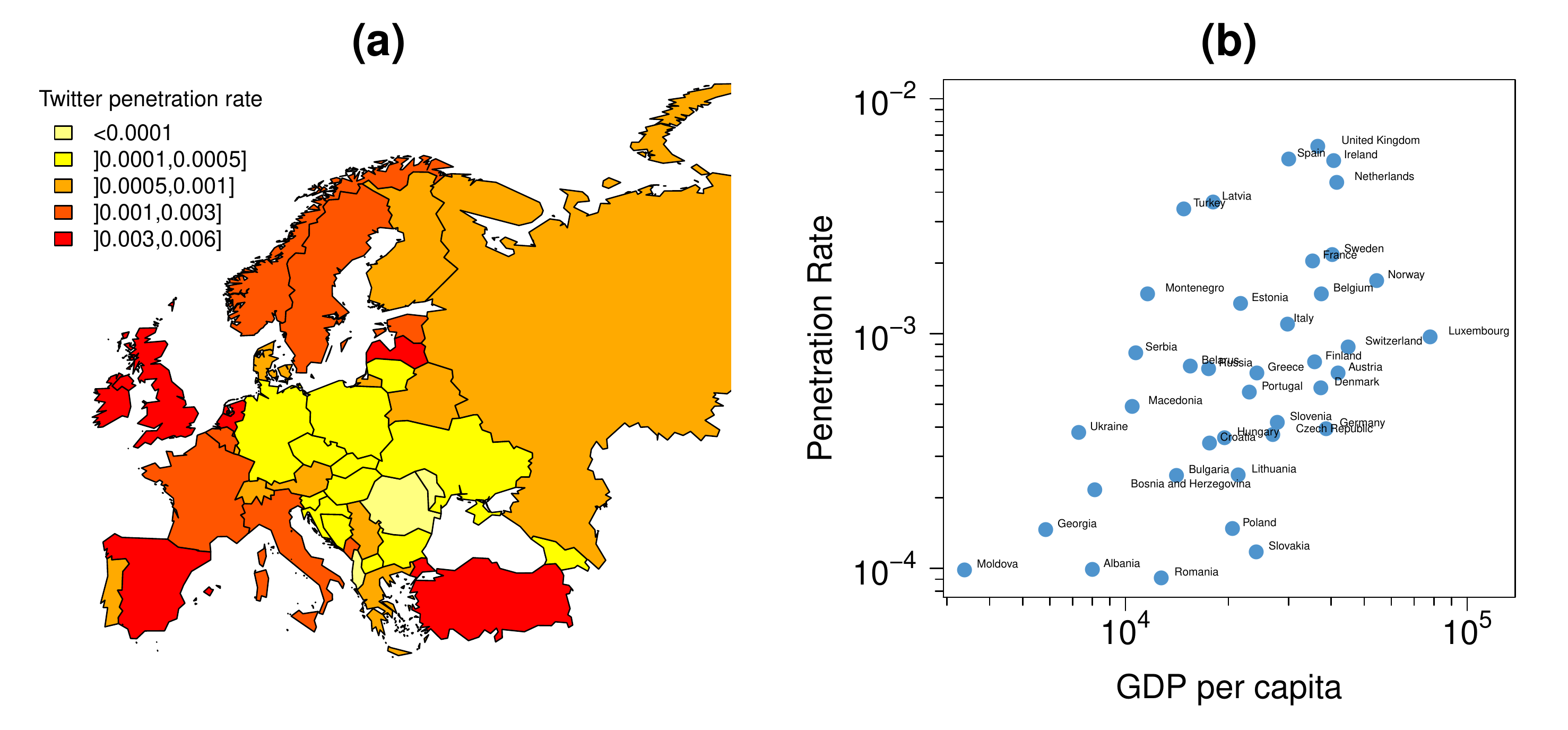}
\caption{(a) Geo-located Twitter penetration rate across Europe country by country defined as the ratio between the number of users emitting geo-located tweets in our database and the country population. (b) Penetration rate as a function of the Gross Domestic Product (GDP) per capita. The figures for the GDP were obtained from the web of the International Monetary Fund \cite{gdp}, correspond to the year $2012$ and are expressed in  US dollars. \label{PR}}
\end{center}
\end{figure*} 

The highway (both directions) and the railway European networks were extracted from OpenStreetMap \cite{OSM} (see Figure \ref{map}b and \ref{map}c for maps of roads and railways, respectively). A close look at the three maps reveals that while tweets concentrate in cities, there is a number of tweets following the main roads and train lines. In this sense, even roads that go through relatively low population areas can be clearly discerned such as those on Russia connecting the main cities, the area of Monegros in Spain, North of Zaragoza or the main roads in the center of France (see the maps country by country in appendix). Here we analyze in detail the statistics of the tweets posted on the roads and railways and discuss the possibility that they are a proxy for traffic and cultural differences. It is important to stress that we considered only the main highways (motorways and international primary roads), not rural roads, while for railways we considered all the main lines (standard gauge of the considered country). The European highways and railways that we consider have a total length of $274,365$ kilometers and $451,475$ kilometers, respectively, which have been divided into segments of $10$ kilometers each. The histograms of total lengths by country of highways (panel (a)) and railways (panel (b)) are plotted in Figure \ref{RankRoad}. Russia, Spain, Germany, France and Turkey represent $50\%$ of the highways total length in Europe. While, for the railway, Russia and Germany represent $35\% $ of the total length. Figure \ref{RankRoad}c shows that most of European countries have a railway network larger than the highway network except for Turkey, Norway, Greece, Spain, Portugal and Finland. In particular, Turkey has a highway system three times larger than the railway network.

\subsection{Identify the tweets on the road}

To identify the tweets on the road/rail, we have considered all the tweets geo-located less than $20$ meters away from a highway (both directions) or a railway. Then, each tweet on the road/rail is associated with the closest segment of road/rail. Using this information, we can compute the percentage of road and rail segments covered by the tweets (hereafter called highway coverage and railway coverage). A segment of road or rail is covered by tweets if there is at least one tweet associated with this segment.  

\subsection{Ethics statement} 

The data analyzed are publicly available as they come from public online sites (Twitter \cite{API} and OpenStreetMap \cite{OSM}). Furthermore, the Twitter data have been anonymized and aggregated before the analysis that has been performed in accordance with all local data protection laws.

\section{RESULTS}

\subsection{General features}

\subsubsection{Penetration rate}

To evaluate the representativeness of the sample across European countries, the Twitter penetration rate, defined as the ratio between the number of Twitter users and the number of inhabitants of each country, is plotted in Figure \ref{PR}a. This ratio is not distributed uniformly across European countries. The penetration rate is lower in countries of central Europe. It has been shown in previous studies \cite{Mocanu2013,Hawelka2013} that the Gross Domestic Product (GDP) per capita (an indicator of the economic performance of a country) is positively correlated with the penetration rate at a world-wide scale. Figure \ref{PR}b shows the penetration rate as a function of the GDP per capita in European countries. No clear correlation is observed in this case. This fact does not conflict with the previous results since our analysis is restricted to Europe and as shown in \cite{Mocanu2013}, in this relationship, countries from different continents cluster together. This means that a global positive correlation appears if countries from all continents are considered but it is not necessarily significant when the focus is set instead on a particular area of the world.

\begin{figure*}
 \begin{center}
\includegraphics[width=\linewidth]{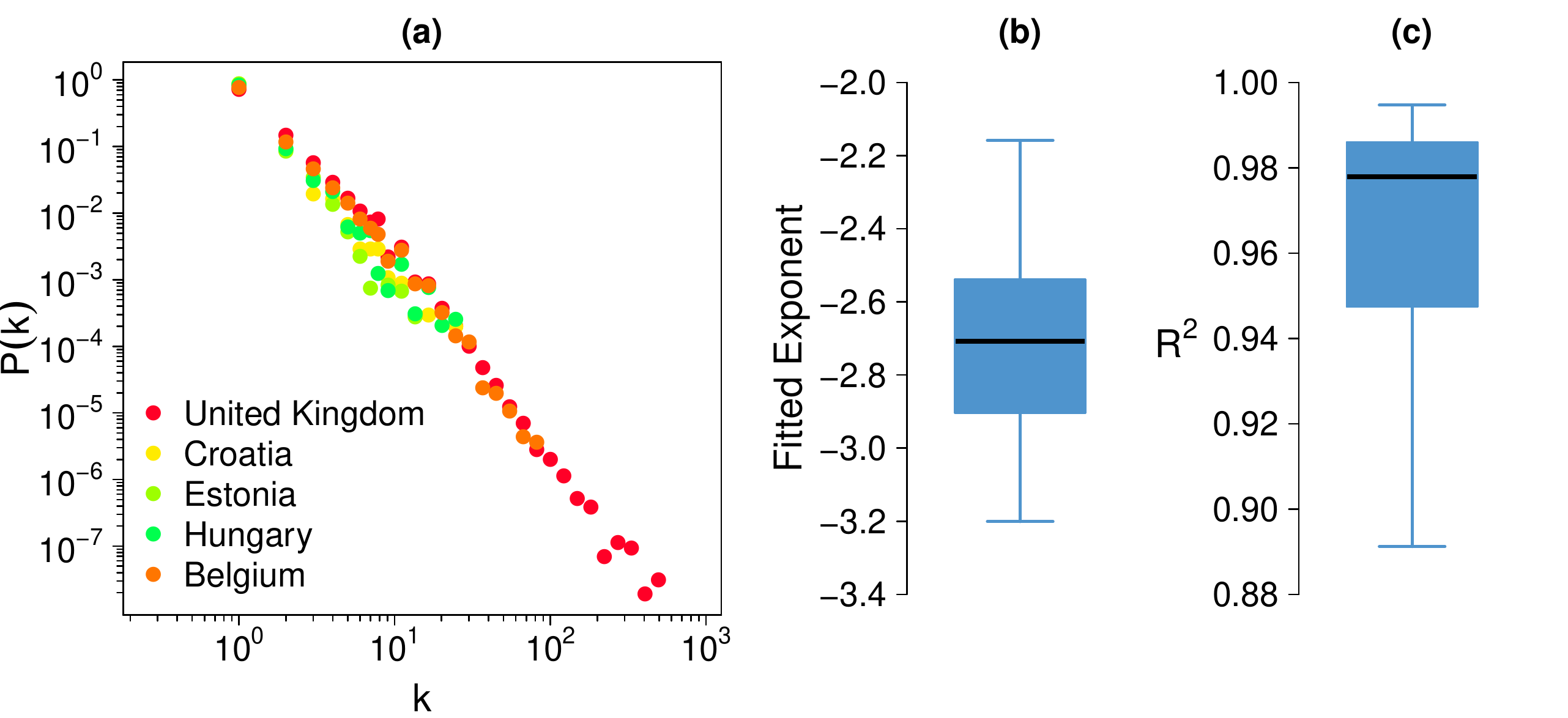}
\caption{(a) Probability distribution of number of ties of an individual in the social network of $5$ countries drawn at random among the 39 case studies. (b) Box plot of the $39$ fitted exponent values. (c) Box plot of the $R^2$ values. The box plot is composed of the minimum value, the lower hinge, the median, the upper hinge and the maximum value.\label{Degree}}
\end{center}
\end{figure*}

\subsubsection{Social network}

The penetration rate of geo-located tweets is different across European countries and does not show a clear relation to the GDP per capita of each country. There are several factors that can contribute to this diversity such as the facility of access or prices of the mobile data providers. In addition, generic cultural differences when facing a delicate issue from the privacy perspective such as declaring the precise location in posted messages can be also present.  One can then naturally wonder whether these differences extend to other aspects of the use of Twitter or are constraint to geographical issues. One obvious question to explore is the structure of the social network formed by the interactions between users. We extract interaction networks by establishing the users as nodes and connecting a pair of them when they have interchanged a reply. Replies are specific messages in Twitter designed to answer the tweets of a particular user. It can be seen as a direct conversation between two users and as shown in \cite{Grabowicz2012} (and references therein) can be related to more intense social relations. A network per country was obtained by assigning to each user the country from which most of his or her geo-located tweets are posted.

Figure \ref{Degree}a shows the distribution of the social network's degree (number of connections per user) of $5$ countries (Belgium, Croatia, Estonia, Hungary and the UK) drawn at random among the $39$ considered. The slope of these $5$ distributions are very similar and can be fitted using a power-law distribution. More systematically, in Figure \ref{Degree}b and Figure \ref{Degree}c we have respectively plotted the box plot of the fitted exponent values obtained for the $39$ countries and the box plot of the R$^2$ associated with these fits. All the $39$ networks have very similar degree distributions, although they show a different maximum degree as a result of the diverse network sizes. These networks are sparse due to the fact that we are keeping only users if they post geo-located tweets and connections only if a reply between two users have taken place. Still and beyond the degree distribution, other topological features such as the average node clustering seems to be quite similar across Europe laying between $0.02$ and $0.04$ for the most populated countries (where we have more data for the network).

\subsection{Highway and railway coverage}

\subsubsection{Dependence between coverage and penetration rate}

\begin{figure*}
\begin{center}
\includegraphics[width=\linewidth]{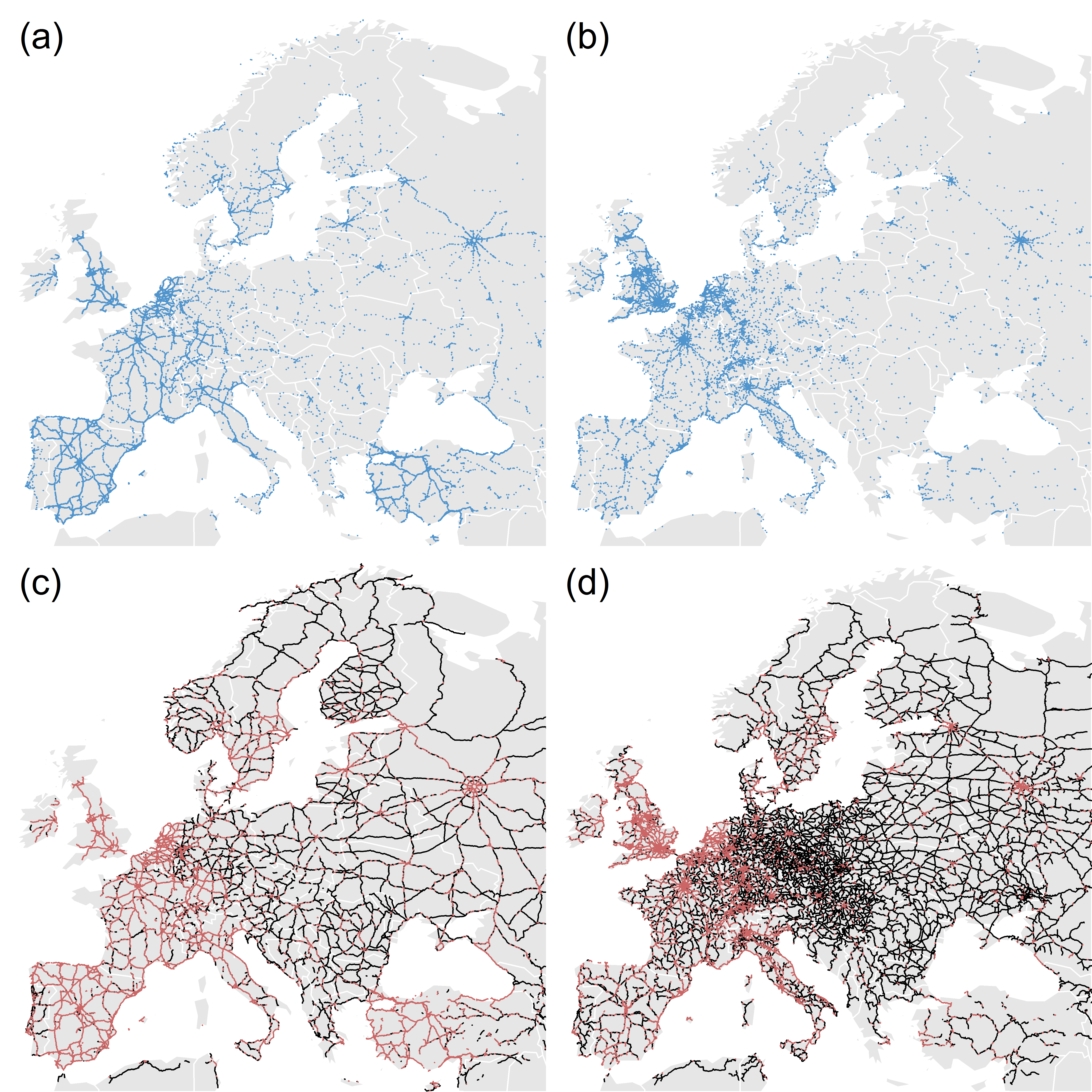}
\caption{(a)-(b) Locations of the geo-located tweets on the road (a) and rail (b). (c)-(d) Segments of road (c) and rail (d) covered by the tweets. The red segments represent the segments covered by the tweets. \label{tweeton}}
\end{center}
\end{figure*}

The percentage of segments (i.e., km) covered by the tweets in Europe is $39\%$ for the highway and $24\% $for the railway. The highway coverage is better than the railway coverage probably because the number of passenger-kilometers per year, which is the number of passengers transported per year times kilometers traveled, on the rail network is lower. However, the coverage is very different according to the country. Indeed, in Figure \ref{tweeton} we can observe that western European countries have a better coverage than countries of eastern Europe except Turkey and, to a lesser extent, Russia. 

\begin{figure*}
\begin{center}
\includegraphics[width=\linewidth]{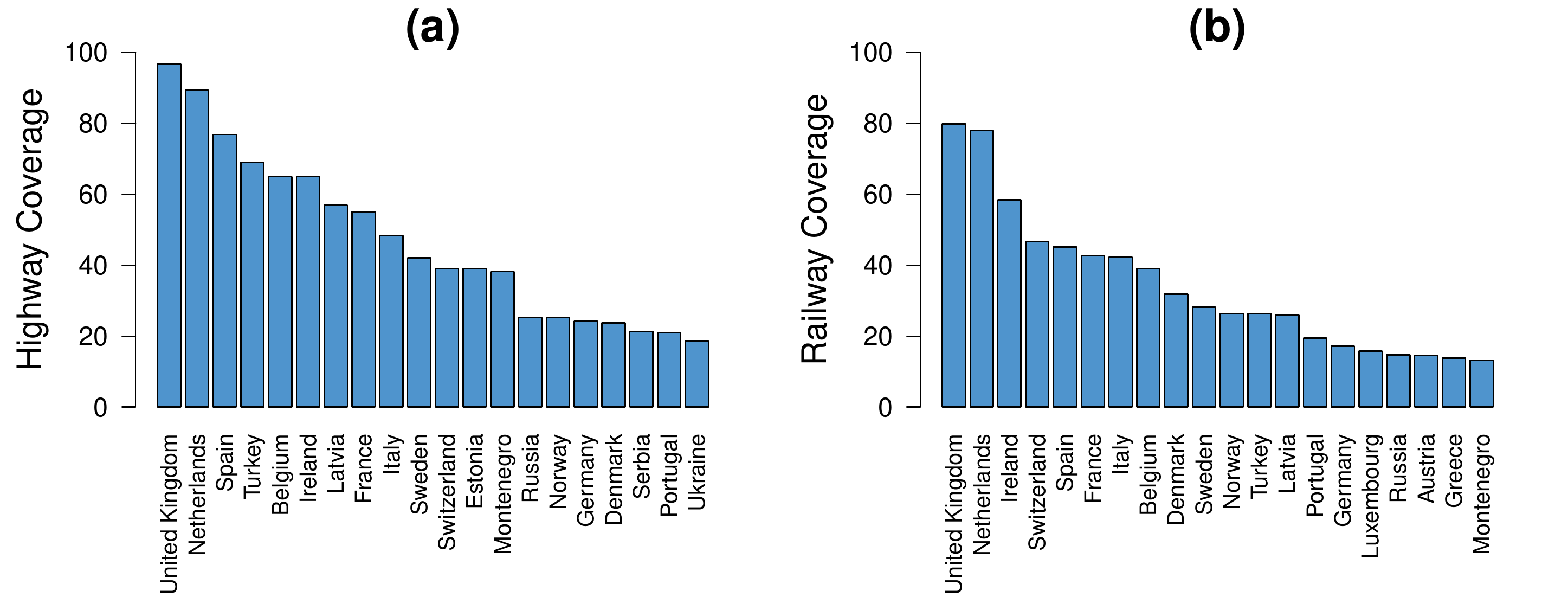}
\caption{Top 20 of the countries ranked by highway coverage (a) and railway coverage (b). \label{Rank}}
\end{center}
\end{figure*} 

\begin{figure*}
\begin{center}
\includegraphics[scale=0.5]{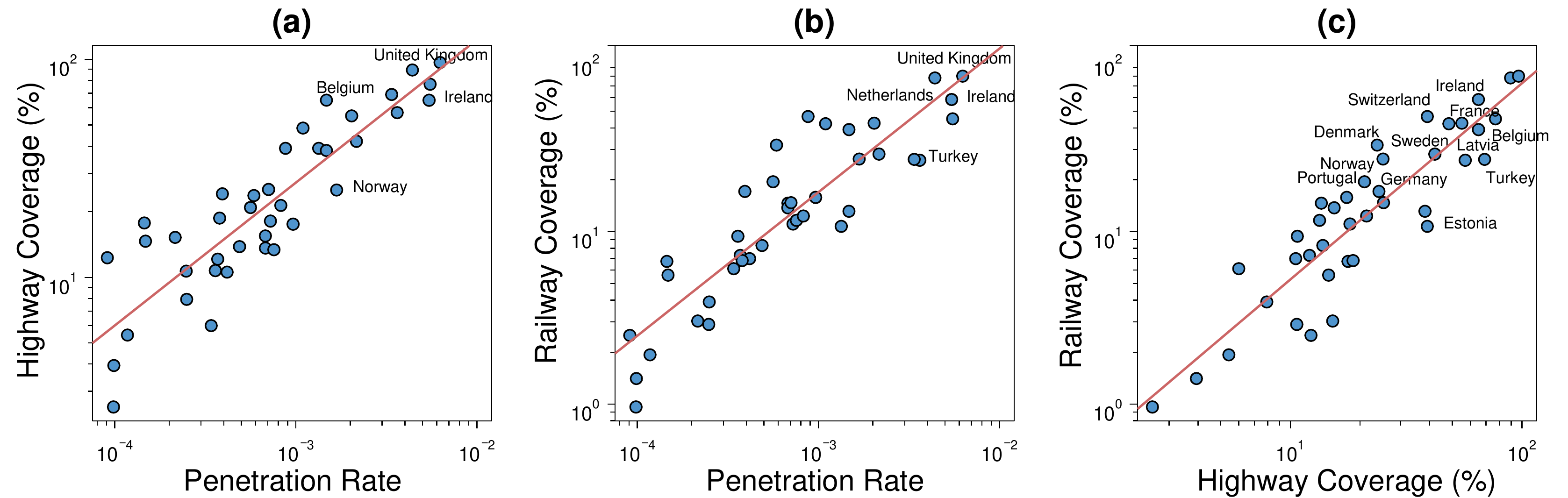}
\caption{(a) Highway coverage as a function of the penetration rate. (b) Railway coverage as a function of the penetration rate. (c) Railway coverage as a function of the highway coverage. Red lines are linear regression curves applied to the log-log plots. This mean that the slope corresponds to the exponents of power-law relationships. The slope in (a) is around $0.6$, $0.8$ in (b) and $1$ (a linear relation) in (c). \label{PRHRHR}}
\end{center}
\end{figure*}

Figure \ref{Rank} shows the top $20$ European countries ranked by highway coverage (Figure \ref{Rank}a) and railway coverage (Figure \ref{Rank}b). The two countries with the best highway and railway coverages are the United Kingdom and the Netherlands. The tweets cover $97\% $ of the highway system in UK and $89\%$ in Netherlands. On the other hand, the tweets cover up to $80\%$ of the railway network in the UK and $78\%$ in Netherlands. Inversely, the country with the lowest coverage is Moldavia with a highway coverage of $2.5\%$ and a railway coverage of $1\%$. The first factor to take into account to understand such differences is the penetration rate. In fact, as it can be observed in Figure \ref{PRHRHR}a and Figure \ref{PRHRHR}b, as a general trend, the coverage of both highway and railway networks is positively correlated with the penetration rate. And, as a consequence, a positive correlation can also be observed between the highway coverage and the railway coverage (Figure \ref{PRHRHR}c). However, these relationships are characterized by a high dispersion around the regression curve. Note that the dispersion is higher than what it can look in a first impression because the scales of the plots of Figure \ref{PRHRHR} are logarithmic. For the two first relationships the mean absolute error is around $7.5 \%$ and for the third one the mean absolute error is around $6.5\%$. This implies that divergences on the geo-located Twitter penetration does not fully explain the coverage differences between the European countries.

\subsubsection{Differences across European countries}

\begin{figure*}
 \begin{center}
\includegraphics[scale=0.65]{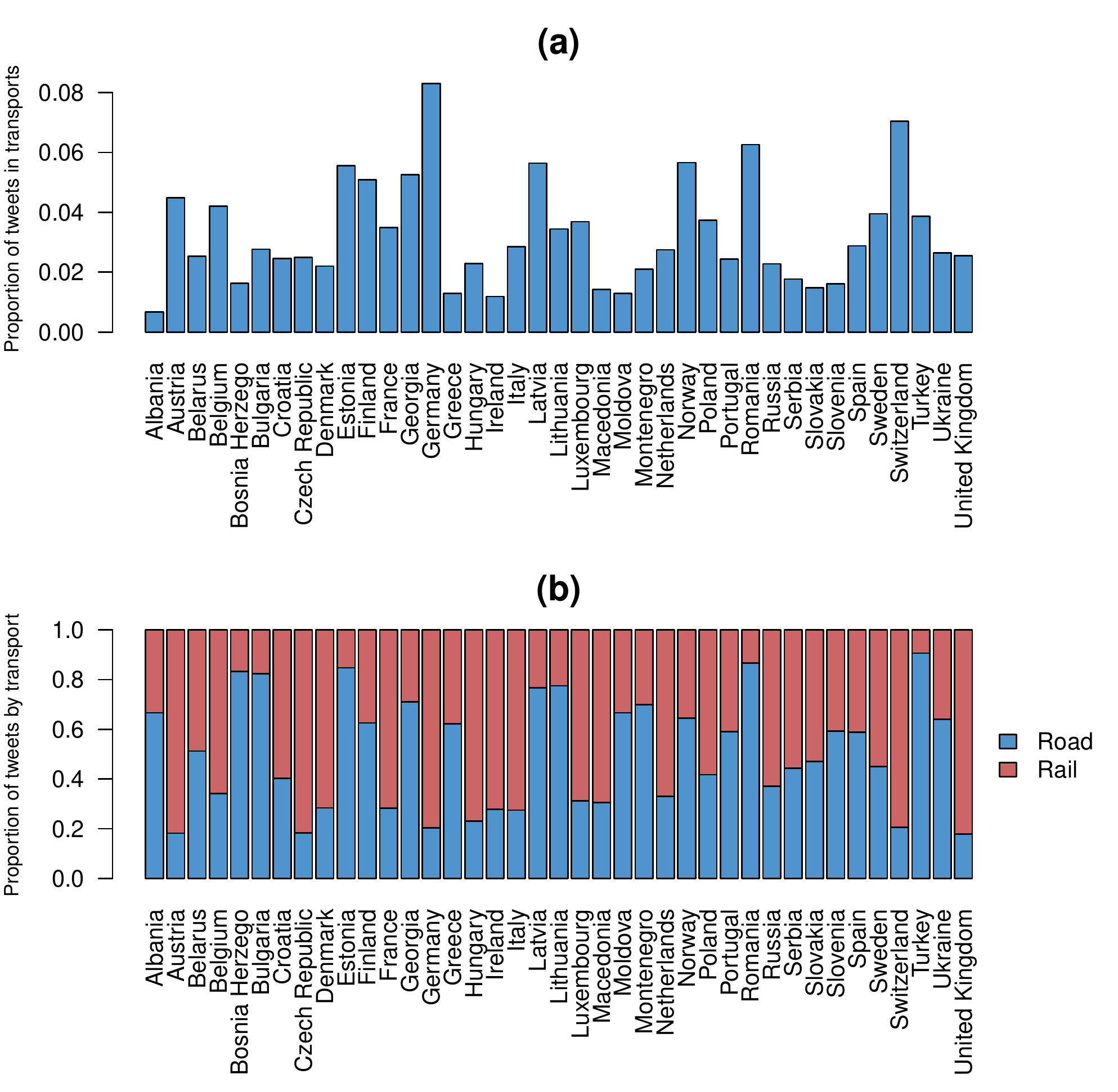}
\caption{(a) Proportion of tweets on the highway or railway networks by European countries. (b) Proportion of tweets according to the transport network by European countries. The blue color represents the tweets on the road. The red color represents the tweets on the rail. The proportions have been normalized in order to obtain a total proportion of tweets in transport equal to $1$. \label{TTOTR}}
\end{center}
\end{figure*}

Disparity in coverage between countries can neither be satisfactorily explained by differences in fares or accessibility to mobile data technology. For example, two countries as France and Spain are similar in terms of highway infrastructure, mobile phone data fares and accessibility, but the geo-located Twitter penetration rates are very different as also are their highway coverage $77\%$ in Spain and $55\%$ in France. Besides penetration rates, divergences in coverage might be the product of cultural differences among European countries when using Twitter in transportation. As it can be observed in Figure \ref{TTOTR}a, the proportion of tweets geo-located on the highway or railway networks is very different from country to country. In the following, we focus on three examples of countries with similar characteristics in the sense of penetration rates but displaying significant differences in transport network coverage.\\

\textbf{Ireland and United Kingdom} The most explicit example of the impact of cultural differences on the way people tweet in transports could be given by the Ireland and United Kingdom case studies. Indeed, these two countries have very similar penetration rates but UK has a proportion of tweets in transports more than two times higher than Ireland. Moreover, both highway and railway coverages are one and a half times higher in UK than in Ireland.\\

\textbf{Turkey and Netherlands} Turkey and Netherlands, which have similar penetration rate, are also an interesting example illustrating how cultural and economical differences may influence coverage. Despite the fact that they both have a high highway coverage, Netherlands has a railway coverage three times higher than Turkey. Different economic levels of train and car travelers in Turkey could be, for instance, an explanation for this.\\ 

\textbf{Belgium and Norway} For countries having similar penetration rate, the higher the proportion of tweets in transports, the better the coverage. However, some exceptions exist, for example, Norway has a proportion of tweets in transports higher than Belgium but, inversely, Belgium has a highway coverage three times higher than Norway. Given the very extensive highway system of Norway, some of the segments, especially on the North, can have very low traffic, which could be the origin of this difference.\\ 

\begin{table*}
	\caption{Difference between the percentage of rail passenger-kilometers and difference between the railway coverages for each pair of countries having a highway coverage higher than 20\% and an absolute different between their highway coverages lower than 5\%.}
	\label{table}
		\begin{center}
			\begin{tabular}{>{\centering}m{4cm} >{\centering}m{4cm} m{4cm}<{\centering}}
				\hline
				 \textbf{Pair of countries} & \textbf{Difference between the percentage of rail passenger-kilometers} & \textbf{Difference between the railway coverages}\\
				\hline
Belgium-Turkey	&	0.13	&	4.8	\\
Ireland-Belgium	&	0.19	&	-4.3	\\
Ireland-Turkey	&	0.32	&	0.5	\\
France-Latvia	&	0.17	&	5.2	\\
Switzerland-Sweden	&	0.18	&	8.1	\\
Sweden-Estonia	&	0.17	&	7.5	\\
Switzerland-Estonia	&	0.36	&	15.6	\\
Norway-Germany	&	0.09	&	-3.6	\\
Danmark-Norway	&	0.05	&	4.5	\\
Norway-Portugal	&	0.07	&	0.2	\\
Danmark-Germany	&	0.15	&	0.9	\\
Portugal-Germany	&	0.02	&	-3.8	\\
Danmark-Portugal	&	0.12	&	4.7	\\
				\hline
	  	\end{tabular}
	  \end{center}
\end{table*}

\begin{figure}
\begin{center}
\includegraphics[scale=0.5]{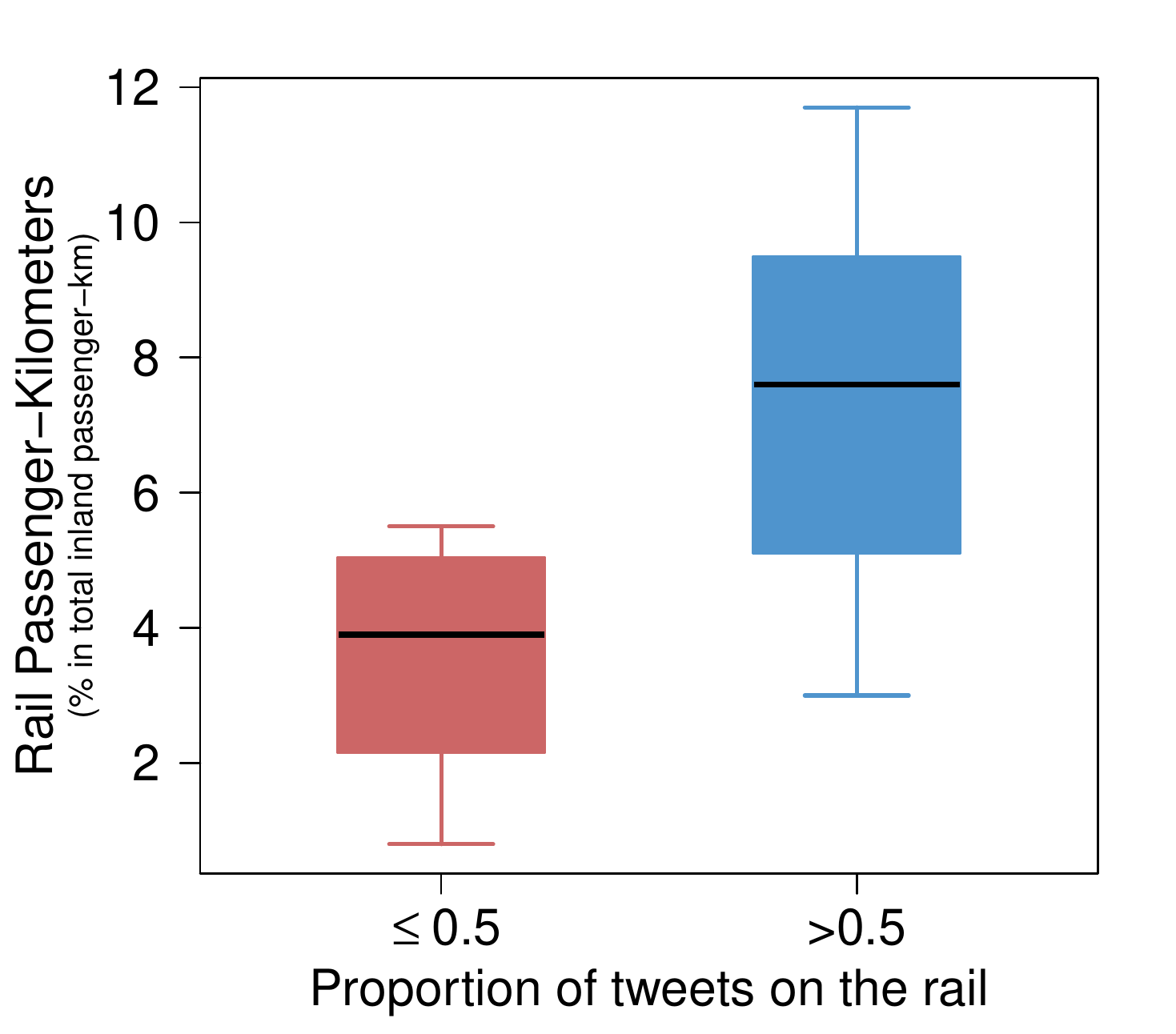}
\caption{Box plots of the distribution of rail passenger-kilometers expressed in percentage of total inland passenger-kilometers according to the proportion of tweets on the rail among the tweets on the road and rail. The red color for countries having a proportion of tweets on the rail less than 0.5. The blue color for countries having a proportion of tweets on the rail higher than 0.5. \label{RPK}}
\end{center}
\end{figure}

In general, the distribution of tweets according to the transport network is also very different from country to country (Figure \ref{TTOTR}b) but also region by region. For example, countries from North and Central Europe have a higher proportion of tweets on the road than tweets on the rail than others European countries. This is probably due to difference regarding the transport mode preference among European countries. To check this assumption, we studied the distribution of rail passenger-kilometers in $2011$ \cite{Eurostat} according to the proportion of tweets on the rail. Figure \ref{RPK} shows box plots of the distribution of rail passenger-kilometers expressed in percentage of total inland passenger-kilometers according to the proportion of tweets on the rail among the tweets on the road and rail. Globally, the number of rail passenger-kilometers is lower for countries having a low proportion of tweets on the rail, which confirms our assumption.

In the same way, the distribution of rail passenger-kilometers in $2011$ can be used to understand why two countries having the same highway coverage might have very different railway coverages. For example, Switzerland and Estonia have the same highway coverage with about $40\%$ of road segments covered by the tweets but the railway coverage is very different, with about $47\%$ of rail segments covered in Switzerland and $11\%$ in Estonia. This can be explained by the fact that in Switzerland trains accounted for $17.6\%$ of all inland passenger-kilometers in $2011$ (which was the highest value among European countries in that year) and inversely, in Estonia, trains accounted for $2\%$ of all inland passenger-kilometers (one of the lowest in Europe). More systematically, for each pair of countries having similar highway coverages, we compared the difference between railway coverages and the difference between the percentage of rail passenger-kilometers. First, pair of countries having a highway coverage higher than $20\%$ and an absolute different between their highway coverages lower than $5\%$ are selected. Thus, we have selected $13$ pairs of countries with a similar highway coverage. Table \ref{table} displays the difference between the percentage of rail passenger-kilometers and the difference between the railway coverages for these $13$ pairs of countries. In $10$ out of $13$ cases, the differences have the same sign. This fact points towards a possible correlation between traffic levels and tweet coverage.

\subsection{Average Annual Daily Traffic}

To assess more quantitatively this hypothetical relation between the number of vehicles and the number of tweets on the road, we compared the number of tweets and the Average Annual Daily Traffic (AADT) on the highways in United Kingdom in $2012$ \cite{UK} and in France in $2011$ \cite{FR}. The AADT is the total number of vehicle traffic of a highway divided by $365$ days. The number of highway segments for which the AADT was gathered is $877$ in UK and $1974$ in France. The average length of these segments is $3.6$ kilometers in UK and $5.8$ in France. As in the previous analysis, the number of tweets associated with a segment was computed by identifying all the tweets geo-located at less than $20$ meters away from the segment. 

\begin{figure*}
\begin{center}
\includegraphics[scale=0.6]{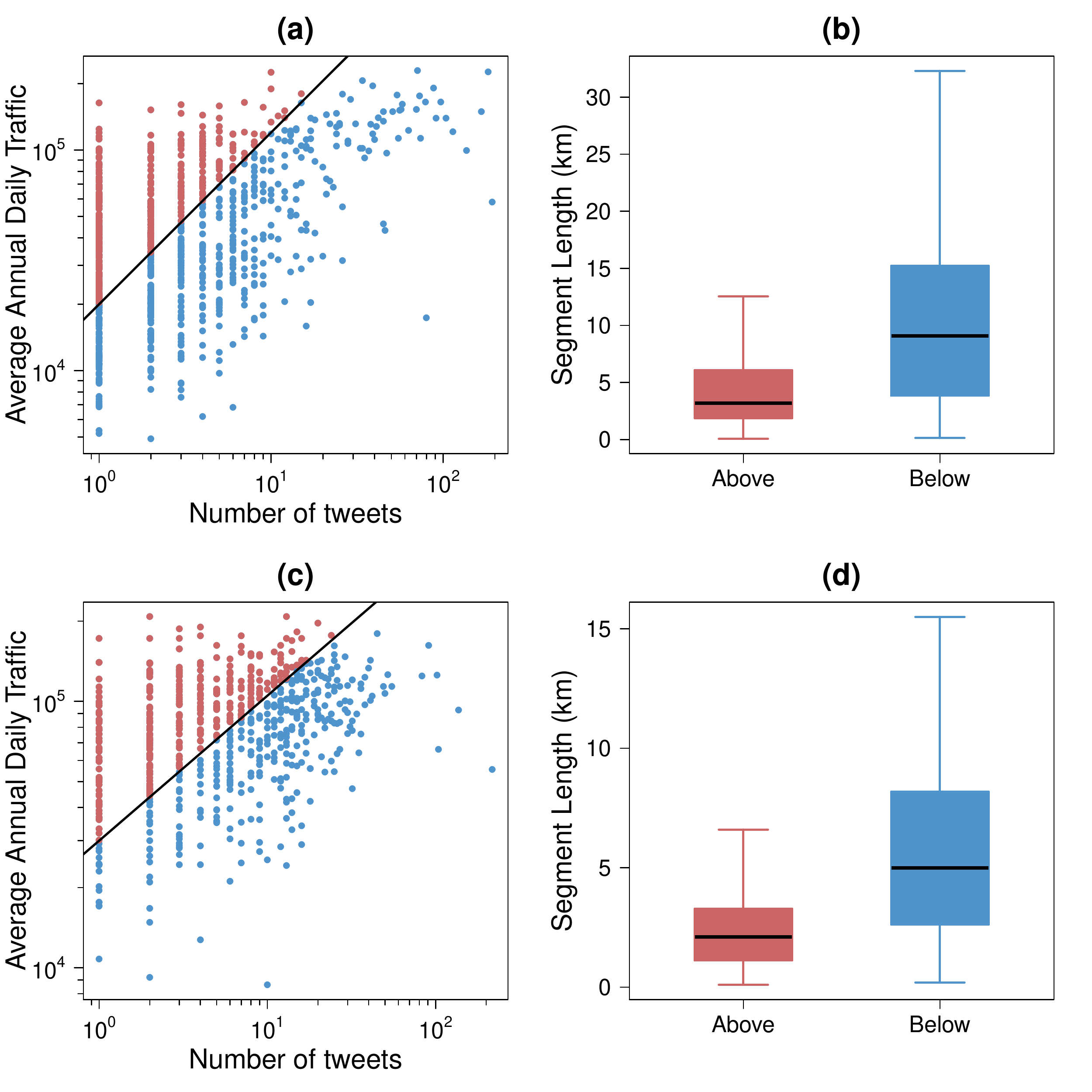}
\caption{(a)-(c) Comparison between the average annual daily traffic and the number of tweets on the highways in France (a) and United Kingdom (b). Points are scatter plots for each highway segments. (b)- (d) Box plots of the highway segment length according to the type of segment in France (b) and United Kingdom (d). The red color for the points above the black line and the blue color for the points below the black line. \label{FRUK}}	
\end{center}
\end{figure*}

Figure \ref{FRUK}a and \ref{FRUK}c shows a comparison between the AADT and the number of tweets on the road for both case studies. There is a positive correlation between the AADT and the number of tweets on the road but the Pearson correlation coefficient values are low, around $0.5$ for the France case study and around $0.3$ for the UK case study. This can be explained by the large number of highway segments having a high AADT but a very low number of tweets. To understand the origin of such disagreement between tweets and traffic, we have divided the segments into two groups: those having a high AADT and a very low number of tweets (red points) and the rest (blue points). These two types of segments have been separated using the black lines in Figure \ref{FRUK}a and \ref{FRUK}c. Figure \ref{FRUK}b and \ref{FRUK}d show the box plots of the highway segment length in kilometer according to the segment type for both case studies. It is interesting to note that the segments having a high AADT and a low number of tweets are globally shorter than the ones of the other group. Indeed, according to the Welch two sample t-test \cite{Welch1951} the average segment length of the first group ($5.2$ km in France and $2.5$ in UK) is significantly lower than the one of the second group ($11.3$ km in France and $5.9$ in UK). Given a similar speed one can assume that the shorter the road segment is, the lower time people have to post a tweet. Other factors that may influence this result is the nature of the segments, rural vs urban, and the congestion levels that can significantly alter the time spent by travelers in the different segments.

\section{DISCUSSION}

In this work, we have investigated the use of Twitter in transport networks in Europe. To do so, we have extracted from a Twitter database containing more than $5$ million geo-located tweets posted from the highway and the railway networks of $39$ European countries. First, we show that the countries have different penetration rates for geo-located tweets with no clear dependence on the economic performance of the country. Our results show, as well, no clear difference between countries in terms of the topological features of the Twitter social network. Dividing the highway and railway systems in segments, we have also studied the coverage of the territory with geo-located tweets. European countries can be ranked according to the highway and railway coverage. The coverages are very different from country to country. Although some of this disparity can be explained by differences in penetration rate or by the use of different transport modalities, a large dispersion in the data still persist. Part of it could be due to cultural differences among European countries regarding the use of geo-located tools. Finally, we explore whether Twitter can be used as a proxy to measure of traffic on highways by comparing the number of tweets and the Average Annual Daily Traffic (AADT) on the highways in United Kingdom and France. We observe a positive correlation between the number of tweets and the AADT. However, the quality of this relationship is reduced due to the short character of some AADT highway segments. We conclude that the number of tweets on the road (train) can be used as a valuable proxy to analyze the preferred transport modality as well as to study traffic congestion provided that the segment length is enough to obtain significant statistics.   

\section{ACKNOWLEDGEMENTS}
Partial financial support has been received from the Spanish Ministry of Economy (MINECO) and FEDER (EU) under projects MODASS (FIS2011-24785) and INTENSE@COSYP (FIS2012-30634),  and from the EU Commission through projects EUNOIA, LASAGNE and INSIGHT. ML acknowledges funding from the Conselleria d'Educaci\'o, Cultura i Universitats of the Government of the Balearic Islands and JJR from the Ram\'on y Cajal program of MINECO.

\makeatletter
\renewcommand{\fnum@figure}{\small\textbf{\figurename~S\thefigure}}
\makeatother
\setcounter{figure}{0}

\setcounter{equation}{0}

\section*{APPENDIX}

\begin{figure*}[ht]
  \begin{center}
		\includegraphics[width=\linewidth]{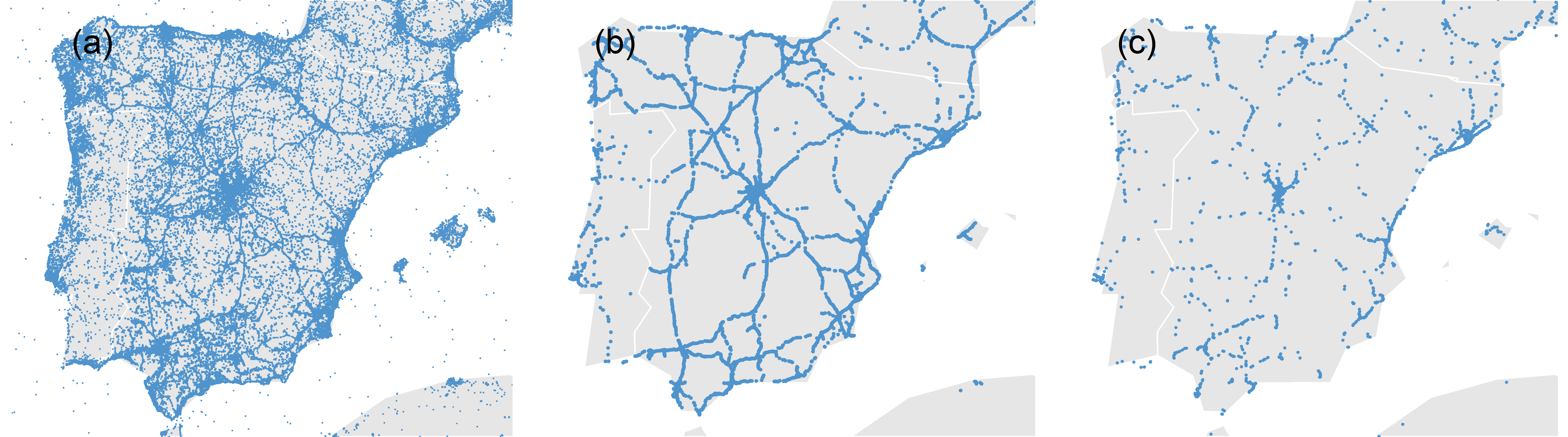}
	\end{center}
	\caption{(a) Locations of the geo-located Tweets (a) on the road (b) and on the rail (c) in Spain. \label{FigS1}}
\end{figure*}

\begin{figure*}[ht]
  \begin{center}
		\includegraphics[width=\linewidth]{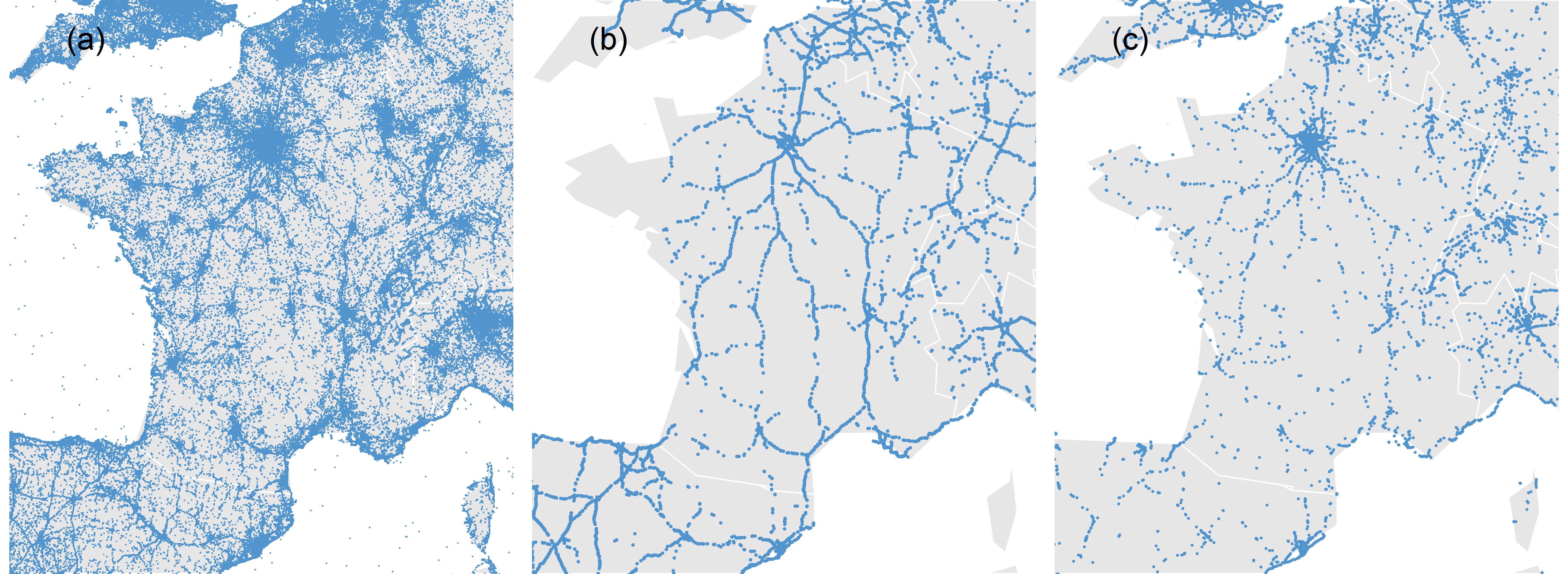}
	\end{center}
	\caption{(a) Locations of the geo-located Tweets (a) on the road (b) and on the rail (c) in France. \label{FigS2}}
\end{figure*}

\begin{figure*}[ht]
  \begin{center}
		\includegraphics[width=\linewidth]{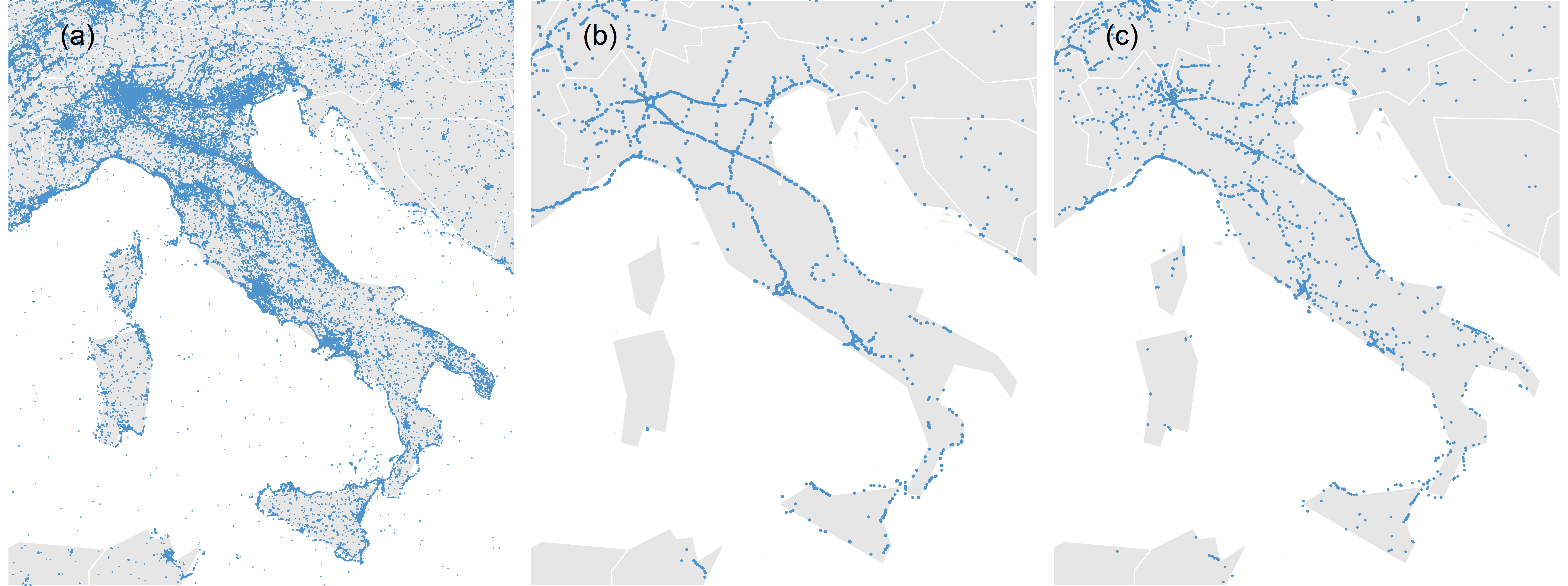}
	\end{center}
	\caption{(a) Locations of the geo-located Tweets (a) on the road (b) and on the rail (c) in Italy. \label{FigS3}}
\end{figure*}

\begin{figure*}[ht]
  \begin{center}
		\includegraphics[width=\linewidth]{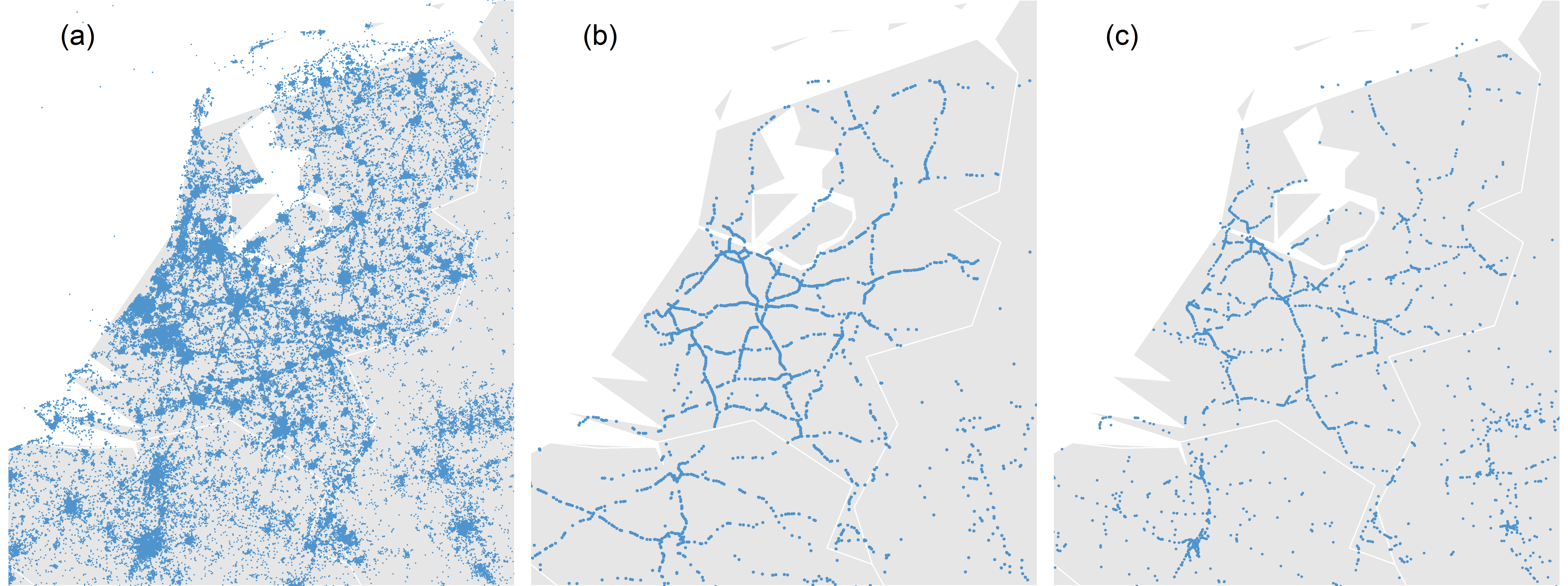}
	\end{center}
	\caption{(a) Locations of the geo-located Tweets (a) on the road (b) and on the rail (c) in Netherlands. \label{FigS4}}
\end{figure*}

\begin{figure*}[ht]
  \begin{center}
		\includegraphics[width=\linewidth]{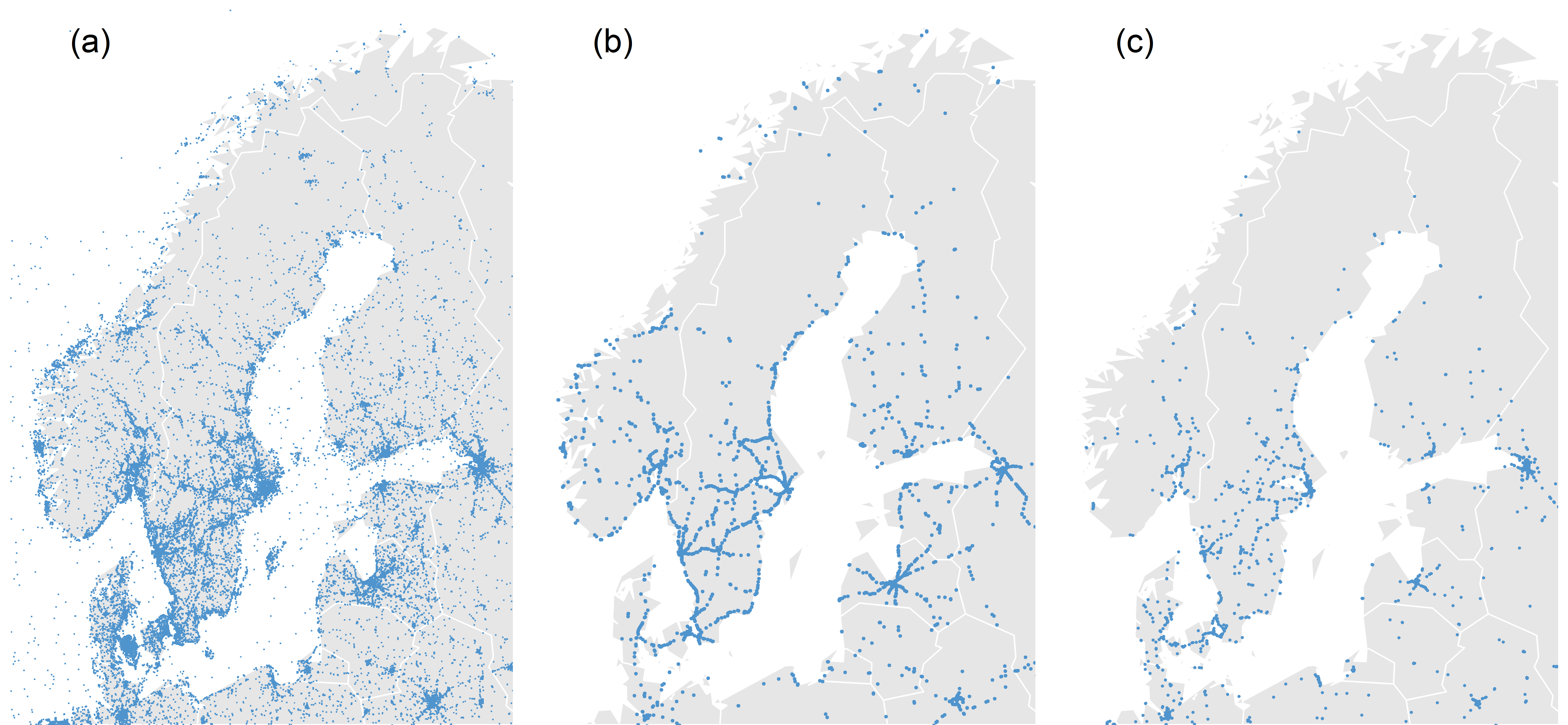}
	\end{center}
	\caption{(a) Locations of the geo-located Tweets (a) on the road (b) and on the rail (c) in Norway. \label{FigS5}}
\end{figure*}

\begin{figure*}[ht]
  \begin{center}
		\includegraphics[width=\linewidth]{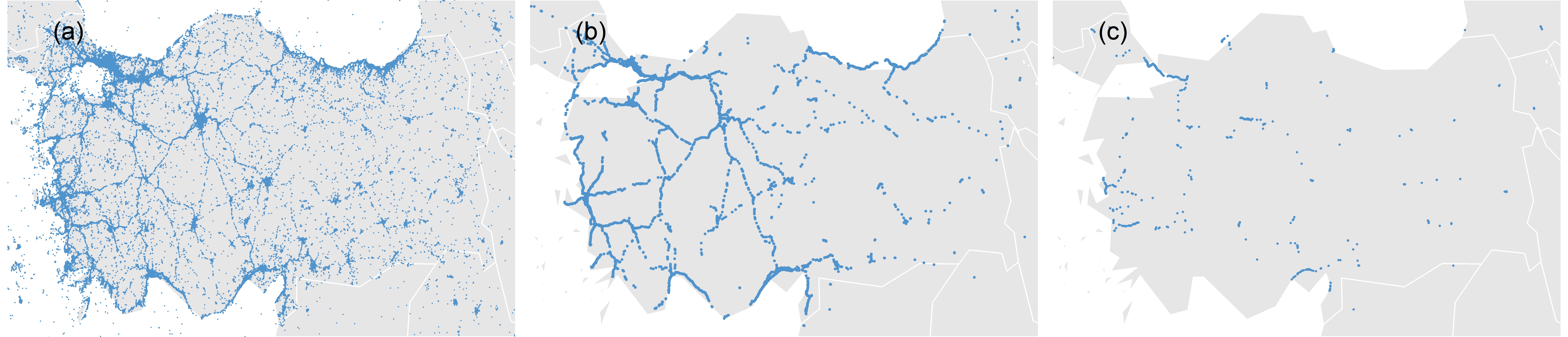}
	\end{center}
	\caption{(a) Locations of the geo-located Tweets (a) on the road (b) and on the rail (c) in Turkey. \label{FigS6}}
\end{figure*}

\begin{figure*}[ht]
  \begin{center}
		\includegraphics[width=\linewidth]{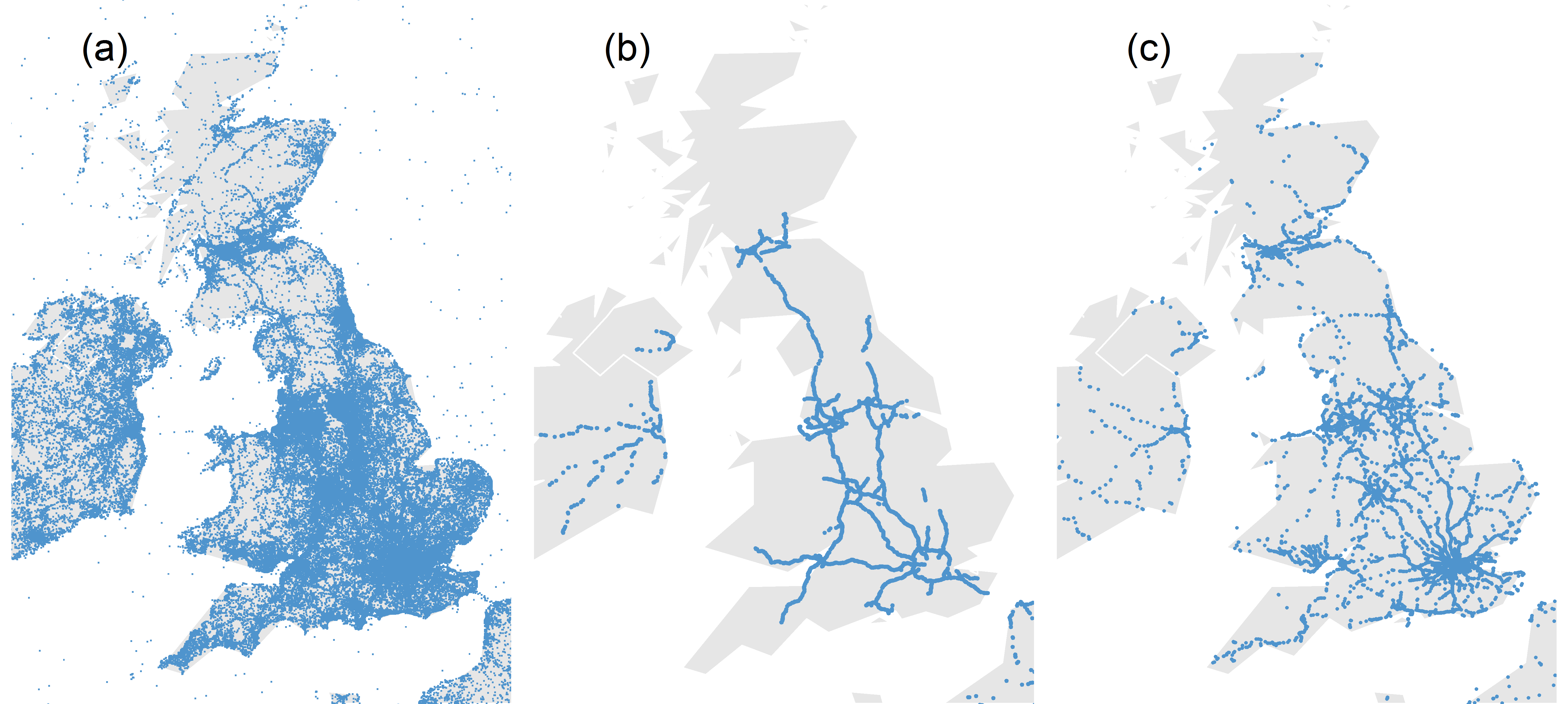}
	\end{center}
	\caption{(a) Locations of the geo-located Tweets (a) on the road (b) and on the rail (c) in United Kingdom. \label{FigS7}}
\end{figure*}

\end{document}